# Periodicity of ~155 days in solar electron fluence

**Partha Chowdhury**

Department of Physics, West Bengal University of
Technology, Kolkata 700 064, W. B., India; partha240@yahoo.co.in

**P. C. Ray**

Department of Mathematics, Government College of
Engineering and Leather Technology, Kolkata 700 098, W. B., India;
raypratap1@yahoo.co.in

**Saibal Ray**

Department of Physics, Barasat Government College,
Kolkata 700 124, North 24 Parganas, W. B., India and Inter-University Centre for
Astronomy & Astrophysics, Post Bag 4, Pune 411 007, India; saibal@iucaa.ernet.in

**Abstract:**
In this paper we have investigated the occurrence rate of high energetic (E>10 MeV) solar electron flares measured by IMP-8 spacecraft of NASA for solar cycle 21 (June, 1976 to August, 1986) first time by three different methods to detect periodicities accurately. Power-spectrum analysis confirms a periodicity ~155 days which is in consistent with the result of Chowdhury and Ray (2006), that ``Rieger periodicity" was operated throughout the cycle 21 and it is independent on the energy of the electron fluxes.

**Key Words:** Solar cycle, flares, particle emission
**Pacs:** 96.60.qd, 96.60.qe, 96.60.Vg

## 1. INTRODUCTION

The Sun often exhibits different periodicities on many different time scales not only in electromagnetic radiation but also in energetic particle events. As a long term periodicity, the 11-year sunspot cycle (Hale cycle) and for short terms, 27 day rotational periods are most prominent and the regime between these extremes of time scales is called the `mid range' (Bai 2003a). During solar cycle 21 a quasi-period of ~154 days was first discovered by Rieger et al. (1984) in solar γ-ray and soft X-ray flares recorded by the Gamma-Ray Spectrometer (GRS) abroad the Solar Maximum Mission (SMM) and subsequently confirmed in microwave burst data (Bogart and Bai 1985) and also in $H_\alpha$ flares (Ichimoto et al. 1985). Afterwards a number of researchers have extensively searched the mid-term quasi-periodicities (one to several months or longer) of different solar flare activities, energetic particle fluxes, sunspot numbers, areas, photospheric magnetic flux or interplanetary magnetic field for solar cycle 21 (Dennis 1985; Delache et al. 1985; Bai & Sturrock 1987; Lean 1990; Dröge et al. 1990; Gabriel et al. 1990; Pap et al. 1990; Kile & Cliver 1991; Oliver et al. 1992, 1998; Cane et al. 1998; Ballester et al. 2002; Krivova & Solanki 2002). Most of these studies indicate a periodicity ranging from 152 to 158 days, but it appears to be dominant especially in the time phase from ~1979 to 1983 corresponding to the solar activity maximum. Besides this other quasi-periods ~129, ~103, ~ 844, ~78 and ~51 days of different solar data during maxima of various cycles (Landscheit 1986; Bai & Sturrock 1991; Sturrock & Bai 1992; Bai 2003a; Atac et al. 2006; Joshi et al. 2006) were reported time to time by different researchers. In a recent paper (Chowdhury & Ray 2006) have studied the periodicities of electron fluence data of two different energy bands (E > 0.6 MeV and E>2 MeV) for cycles ~21 to ~23 and reported ~152 day periodicity for solar cycle 21 for both energy bands.

Recently Forgacs-Dajka and Borkovits (2007) indicated that the mid-term periodicities are manifest in almost all solar data (sunspot numbers, solar flare index, solar radio flux, IMF, proton speed etc.) with the exception of the coronal index and 10.7 cm solar flux. In the present work, we have extended our study for the data of more energetic electron fluence (E>10 MeV) for complete solar cycle 21 (June, 1976 - August, 1986) first time by different power spectrum analysis methods to determine periodicities accurately because they can provide better information on properties of the Sun.

## 2. DATA AND ANALYSIS METHODS

The data employed in this study are the daily-averaged electron fluences observed by IMP-8 spacecraft of NASA in energy range >10 MeV and are residual after the background is subtracted. This electron flux data for period 1 June, 1976 to 31 August, 1986 were taken from OMNI database compiled by the National Space Science Data Centre (NSSDC) and available at the website
http://nssdc.gsfc.nasa.gov/space/imp-8.html

The daily electron flux data were taken in regular manner and the small data gaps found present have been filled in by using an interpolation technique. To investigate the occurrence rate of peak electron flux accurately we have adopted three different spectral decomposition techniques, viz., (1) the fast-Fourier transformation (FFT), (2) the maximum entropy method (MEM) and (3) the Lomb-Scargle periodogram (LSP). Finally have made a comparative study among the results of all these spectral analysis methods.

### 2.1 Fast-Fourier Transformation (FFT)
To trace out periodicities of electron flux data of IMP satellite, we first have applied the conventional Fourier Power Spectral Analysis technique. The frequency of observation was one datum per day and the Nyquist critical frequency was 0.5 per day.

The fast-Fourier transform is derived from the discrete-Fourier transform to reduce the computational time considerably and at the same time accuracy of the output obtained from FFT are within tolerable limits.

*2.2 Maximum entropy method (MEM)*

The results of FFT are replicated by MEM which tries to avoid the limited resolution and power `leaking', due to windowing of data, present in the former method. Burg (1975) developed this new form of spectral variance analysis which belongs to the class of methods that fits a statistical model to the data and it shows higher resolutions, especially at low frequencies producing narrow peaks.

The Winer-Khinchin theorem states that the Fourier transform of autocorrelation of any signal is equal to the power spectrum. Van den Bos (1971) has shown that the parameters of a maximum entropy spectral estimation are equivalent to the `ones' in the auto-regressive (AR) model of a random process in real domain. To compute the power spectra by MEM we have applied the algorithm proposed by Press et. al. (2001) (details of the algorithm and other technique of MEM is available in the book Numerical Recipes in C, Cambridge university press). In order to justify the validity of the peaks found from the Fourier analysis and Entropy method, the confidence limits (also known as fiducial limits) for different peaks obtained from the power spectra were calculated (Haber & Runyon 1969; Das et al. 1999, 2003; Chowdhury & Ray 2006). All these confidence levels are above 99%. In effect we are attempting to determine the interval within which any hypothesis concerning the periodicity of a certain solar event might be considered tenable and outside which that hypothesis would be considered untenable. The confidence limits (CL) are evaluated by generating a sample of 100 data points equally on both sides of a particular peak. This method is repeated for all the peaks under consideration within a spectrum. The peak that is sharp gives the minimum value of standard error ($SE_m$) and the interval between the CL is reduced, increasing the probability of its being near $T_{ave}$, where $T_{ave}$ is the mean value of
the time period of the sample data points.

A detailed mathematical formalism has been provided in the article by Chowdhury and Ray (2006) and hence here we shall only mention some of the essential definitions as appeared above. The confidence interval or limit provides the lower and upper limits to which the population parameter has a high probability of being included. The population parameter standard deviation $\sigma$ can be calculated from the following formula

$$\sigma = \{\sum T^2 / (N-1) - (\sum T)^2 / N (N-1)\}^{0.5}.$$

The standard error (S.E) is the standard deviation of the sample mean (from sampling distribution) is estimated as:

$$SE_m = \sigma / (N)^{0.5}.$$

The confidence limits (C.L) for 99% confidence can be computed as
$$C.L = T_{ave} \pm (2.58) (SE_m).$$

*2.3 The Lomb-Scargle periodogram (LSP)*
As mentioned earlier that using an interpolation technique we fill small gaps in time series and later to confirm the periodicities obtained by FFT and MEM technique, the parent (non-interpolated) data sets were analyzed by Lomb-Scargle method by calculating the Scargle normalized periodogram (Scargle 1982) following the procedures of Horne and Baliunas (1986). This method successfully handles time series with missing data and provides estimates of the level of significance of the periodogram peaks. The periodogram is determined from the raw daily data, transformed to zero-mean time series, but without smoothing, binning or interpolating (a detailed description of this method is available in the article by Chowdhury and Ray (2006)).

## 3. DISCUSSIONS OF THE RESULTS

*3.1 Criteria for the analysis*
In selecting the periodicity the following criteria were used:

(1) As the time series for complete cycle 21 contains ~3700 data points, only the time periods lying below 1000 days have been considered.

(2) Periodicities ~27 days and its integral multiples were discarded as they coincide with the synodic rotation period of the Sun and its harmonics. So, these periodicities bear the impression of the Sun's synodic rotation and nothing else. After discarding these peaks in this way, the remaining peaks were chosen in the decreasing order of power.

(3) Only the periodicities which are prominently present in all the three different methods have been selected for final consideration.

(4) For periodicity search through FFT and MEM, importance has been given not only to the power of a peak but also on the sharpness of that particular peak. In FFT and MEM the peaks above 2.58σ limits were only considered along their CL values.

*3.2 Analysis of the results*
Figure 1 displays the results of periodicity of energetic electron flux (E>10 MeV) for solar cycle 21 (1976 to 1986) and the results are shown in Table 1.

One can observe that the Table 1 and Figure 1(c) indicate periodicities ~54 days and ~155 days which are common in all the three different methods. However, periodicity ~54 (2 x 27 d) is a subharmonic of solar rotation period and hence discarded. So, the only peak ~155 days is prominently present in all techniques and in Scargle method it is close to 0.1% significance level. Therefore, the occurrence rate of energetic (E>10 MeV) electron flares for solar cycle 21 is ~155 days. Now, it will be useful to compare and discuss our analysis in the light of periodicities reported in other solar activity indicators. In cycle 21, near 155 days periodicity was first detected by Rieger et al. (1984) and later reported in other studies. During the same cycle, it was detected in energetic proton flares (Bai & Cliver 1990; Gabriel et.al. 1990), in ground-based $H_\alpha$ and microwave flares

(Ichimoto et. al. 1985; Bogart & Bai 1985; Kile & Cliver 1991), in interplanetary magnetic field data (Cane et. al. 1998). Lean and Brueckner (1989) noted this periodicity in sunspot blocking function and in 10.7 cm radio flux during solar cycles 19, 20 and 21. Carbonell and Ballester (1990, 1992) suggested that a periodicity ~150 to 160 days had been significant during all solar cycles from 16 to 21. Recently Chowdhury and Ray (2006), in an analysis of electron fluence data for two different energy bands (E>0.6 MeV and E>2 MeV), detected significant periodicity ~152 days for cycle 21. The present analysis confirms that ``Rieger periodicity" was prominently present all over the cycle 21 for electron flares which is in contrast to the opinion of Lean (1990) that this periodicity in sunspot areas occur intermittently in each cycles during the epoch of maximum activity.

## 4. CONCLUSIONS

It is important to note here that no satisfactory theory exists for ~155 days periodicity but several suggestions have been made time to time by several authors. Ichimoto et al. (1985) suggested that it results from strong `magnetized streams' appearing in stack plots of synoptic magnetic charts. But, it has been refuted by Bai and Sturrock (1987), because although the flares occurring in these active regions (Bai 1987a,b) show the periodicity so do those outside of the active region. Lean and Brueckner (1989) linked it with the magnetism of sunspots and suggested that the escape of magnetic field from the Sun is the cause. Mayfield and Lawrence (1985) showed that flare production is correlated with the total magnetic energy of an active region. Recently, Ballester et al. (2002, 2004) have confirmed ~160 days periodicity in the photospheric magnetic flux even for cycle 21 and 23. Based on these analysis researchers (Carbonell et. al. 1990, 1992; Ballester et al. 2002) have proposed that the periodic emergence of magnetic flux, manifested as sunspots, triggers the periodicity in high-energy solar flares, probably by reconnection between old and new magnetic flux. It is interesting to note that Oliver et. al. (1998) detected that during solar cycle 21 there was a perfect time-frequency coincidence between the occurrence of the periodicity in both sunspot areas and high energy flares.

On the other hand, as a possible cause of this periodicity Bai and Cliver (1990) have proposed that this behavior could be simulated with a damped, periodically forced nonlinear oscillator, which shows periodic behavior for some values of the parameters and chaotic behavior for other values. Bai and Sturrock (1991) and Sturrock and Bai (1992) suggest that the Sun contains a `clock' with a period of 25.8 days (later modified to 25.5 days) and the different solar periodicities are subharmonic of that fundamental period. Bai and Sturrock (1993) studied the longitude distribution of major solar flares of the cycles 19 to 22 and explained this hypothetical clock as being an obliquely rotating structure (either magnetic or hydrodynamic) rotating with a period of 25.5 days about an axis tilted by $40^o$ with respect to the solar rotation axis. Interestingly ~155 (which is ~6 x 25.5) days is a integral multiple of the proposed 25.5 days. Thus our result seems to be consistent with that of Bai and Sturrock (1993) made conclusion that ~155 days periodicity is a global phenomenon involving the whole Sun and 25.5 days is the fundamental period of the Sun. It is pertinent to mention here that ~155 days is independent on the flux intensity of electron flares. However, the reason behind the `clock mechanism' is still unknown (Bai 2003b).

Pap et al. (1990) and later Bouwer (1992) suggested that temporary existence of 154 ($\pm$13) days periods in solar activity indices are related to a strong emerging magnetic field. Bai (1987b, 1989) and Sammis et. al. (2000) showed that major solar flare production is associated with super active regions of exceptional longitudinal extent, typically containing very large sunspots having complex magnetic configuration. Furthermore, Bai (2003a) has determined that during solar cycle 21, a double hotspot system having rotation rate 27.41 days was operated in the northern hemisphere of the Sun. The synodic period of 27.41 days corresponds to the sideral period of 25.5 days. Fan et. al. (1994) suggest that due to buoyancy magnetic fields are usually transported from bottom to the upper part of the solar convection zone and this hot spot system may play key role in this movement. The nearly same value of periodicity of electron fluences and photospheric magnetic flux (Ballester et. al. 2002) and other solar magnetic activities make us conclude that electron emission is so intimately connected with the internal solar dynamics that the periodicities of one is reflected in the other. The further study of the other solar activities is needed in order to access the significance of the ~155 days periodicity.

Acknowledgements
The use of IMP-8 observations from NSSDC OMNI data base is gratefully acknowledged. The authors are grateful to Dr. Natalia Papitashvili, Dr. Robert McGuire of NASA /Goddard Space Flight Center and Dr. Clifford Lopate of the University of New Hampshire for providing the necessary data and information in this context. Useful discussion with Dr. I. G. Richardson, NASA Goddard Space Flight Centre, Maryland, U.S.A, about the periodogram analysis is also gratefully acknowledged.

Figure 1. Power spectrum of energetic electron flux data (E>10 MeV) for the solar cycle 21: (a) by FFT ; (b) by MEM ; (c) by LSP.

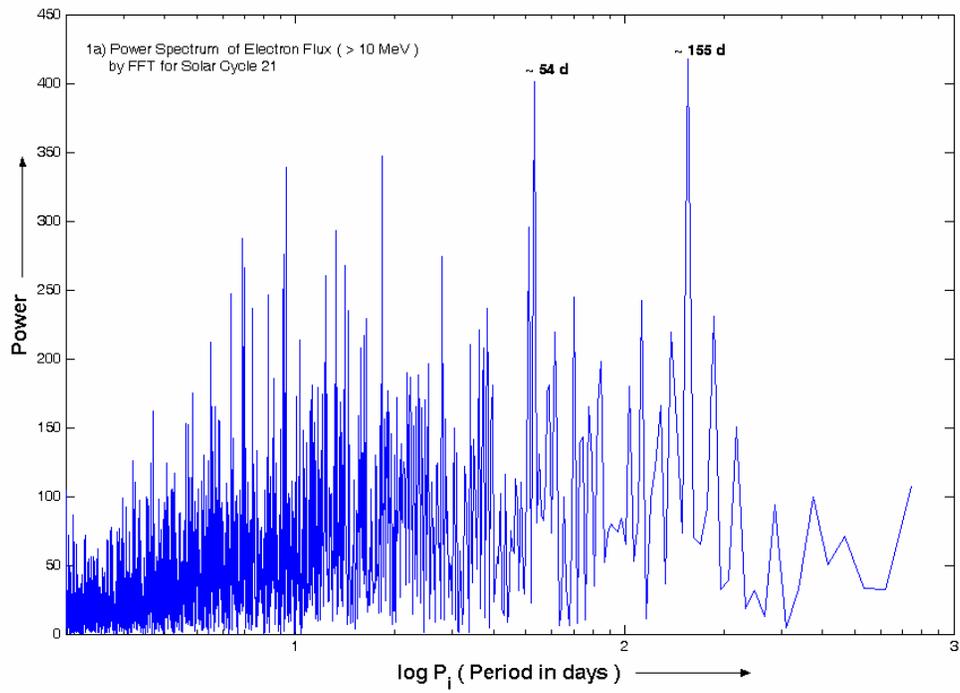

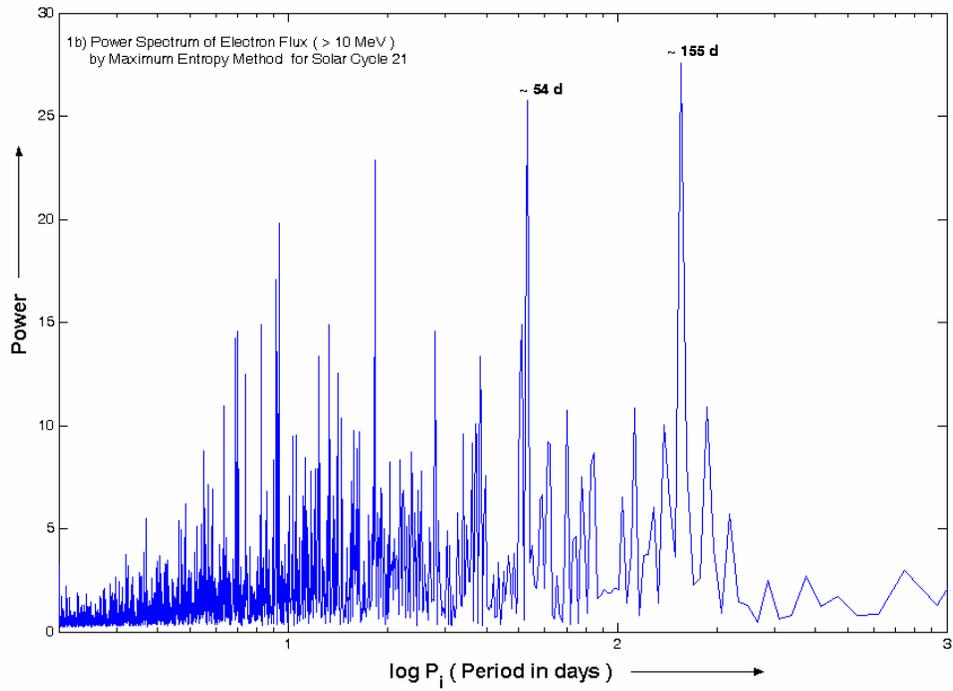

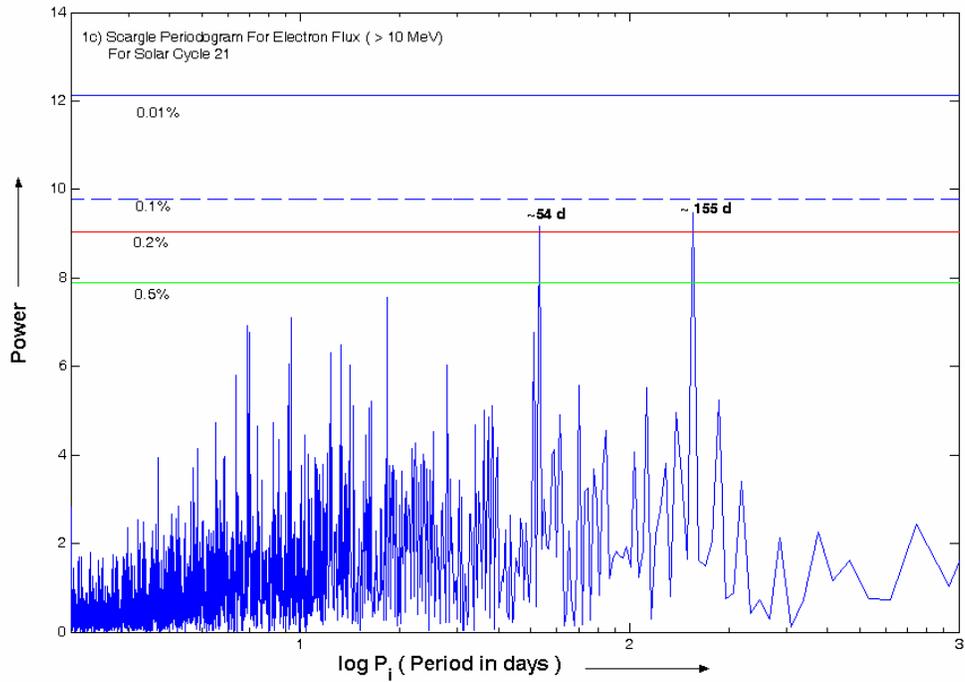

## Table-1

List of periods (in days) detected with their standard errors for confidence limits, in daily electron flux data for solar cycle 21 (1.6.76 – 31.8.86)

| Electron energy | | Identified periods in spectral power peaks | | | | |
|---|---|---|---|---|---|---|
| | | A | B | C | D | E |
| **(a) By FFT** | | | | | | |
| >10 MeV | $T_{ave}$ | 155.67 | 53.26 | 18.27 | 9.34 | 51.06 |
| | $SE_m$ | 0.0188 | 0.0021 | 0.0003 | 0.00006 | 0.0021 |
| **(b) By MEM** | | | | | | |
| >10 MeV | $T_{ave}$ | 155.38 | 53.28 | 18.27 | 9.34 | 9.13 |
| | $SE_m$ | 0.2691 | 0.0306 | 0.0036 | 0.0010 | 0.0009 |